\documentclass[12pt,preprint]{aastex}

\shorttitle{Icarus and 2007 $\mathbf{MK_6}$}
\shortauthors{Ohtsuka et al.}

\begin{document}


\title{Apollo asteroids (1566) Icarus and 2007 $\mathbf{MK_6}$:
     Icarus family members?}


\author{\sc K. Ohtsuka\altaffilmark{1}, \sc H. Arakida\altaffilmark{2}, \sc T. Ito\altaffilmark{3}, \sc T. Kasuga\altaffilmark{4, 3}, \sc J. Watanabe\altaffilmark{3},}
\author{\sc D. Kinoshita\altaffilmark{5}, \sc T. Sekiguchi\altaffilmark{3}, \sc D. J. Asher\altaffilmark{6}, and \sc S. Nakano\altaffilmark{7}}


\altaffiltext{1}{Tokyo Meteor Network, 1--27--5 Daisawa, Setagaya-ku,
             Tokyo 155--0032, JAPAN; ohtsuka@jb3.so-net.ne.jp.}
\altaffiltext{2}{Waseda University, 1--6--1 Nishi-Waseda, Shinjuku-ku,
             Tokyo 169--8050, JAPAN; arakida@aoni.waseda.jp.}
\altaffiltext{3}{National Astronomical Observatory of Japan, 2--21--1 Osawa,
             Mitaka, Tokyo 181-8588, JAPAN; tito@cfca.nao.ac.jp,
             jun.watanabe@nao.ac.jp, t.sekiguchi@nao.ac.jp.}
\altaffiltext{4}{Institute for Astronomy, University of Hawaii, 2680 Woodlawn Drive,
             Honolulu, Hawaii 96822--1897; kasugats@IfA.Hawaii.Edu.}
\altaffiltext{5}{Institute for Astronomy, National Central University, 300 Jhongda Rd.,
             Jhongli, Taoyuan 32001, TAIWAN; kinoshita@astro.ncu.edu.tw.}
\altaffiltext{6}{Armagh Observatory, College Hill, Armagh, BT61 9DG, UK;
             dja@arm.ac.uk.}
\altaffiltext{7}{OAA computing section, 1--3--19 Takenokuchi, Sumoto,
             Hyogo 656--0011, JAPAN; nakano@oaa.gr.jp.}


\begin{abstract}

Although it is more complicated to search for near-Earth object (NEO) families than main belt asteroid (MBA) families, since differential orbital evolution within a NEO family can cause current orbital elements to drastically differ from each other, we have found that Apollo asteroids (1566) Icarus and the newly discovered 2007 $\mathrm{MK_6}$ are almost certainly related. Specifically, their orbital evolutions show a similar profile, time shifted by only $\sim 1000$ yr, based on our time-lag theory. The dynamical relationship between Icarus and 2007 $\mathrm{MK_6}$ along with a possible dust band, the Taurid-Perseid meteor swarm, implies the first detection of an asteroidal NEO family, namely the ``Icarus asteroid family''.

\end{abstract}





\section{Introduction}

Near-Earth Apollo asteroid (1566) Icarus = 1949 MA was discovered by Baade (1949) as a 16th magnitude fast-moving object, on a plate taken using the 48-inch Palomar Schmidt on 1949 June 10, when Icarus approached the Earth to within 0.10 AU, near its descending node. Its orbital parameters were highly unusual: it had smaller semimajor axis ($a$ = 1.08 AU), smaller perihelion distance ($q$ = 0.19 AU), and larger eccentricity ($e$ = 0.83) than any other asteroid known at that time and relatively high inclination ($i$ = $23^\circ$). Icarus remained the record holder in having the smallest $q$ among all asteroids until the discovery of (3200) Phaethon in 1983. Indeed, on account of its small $q$, Icarus was historically of particular interest as to whether the relativistic effects on its orbital motion are detectable (e.g., Shapiro et al. 1971).

In Icarus' subsequent approaches to the Earth in 1968, 1987, and 1996, the following physical data were derived: absolute magnitude $(H) = 15.95$ and $G$-parameter $= -0.04$ (Tedesco 1989); rather high albedo $\sim 0.33$ and diameter $\sim 1$ km (e.g., Harris 1998); fast rotation period $\sim2.273$ hr (e.g., Gehrels et al. 1970; De Angelis 1995) and others \footnote{http://earn.dlr.de/nea/001566.htm}. Especially notable is that Icarus is spectrally classified as a Q-type in Tholen's taxonomy. Q-type asteroids, which generally are spectroscopic analogues of ordinary chondrites (cf.\ McFadden et al. 1984; Hicks et al. 1998; Fevig \& Fink 2007), are regarded as being less space-weathered S-complex asteroids, with a surface age $\le 10$ Myr owing to resurfacing effects (Marchi et al. 2006). Hence Icarus may represent the rather fresh internal structure of a precursor object broken up in recent history. Moreover, with $q \sim 0.19$ AU, the subsolar point on Icarus should reach a temperature of 800 K by solar heating, in which case the solar thermal stress may be a trigger to destroy the asteroid's surface and subsurface. Resurfacing may alternatively be due to the tidal effects of the terrestrial planets (Nesvorn\'y et al. 2005).

For the above reasons, we have expected some ``Icarus Family Members'' (hereafter ``IFM(s)'') to exist in near-Earth space. We have therefore been searching for IFMs based on time-lag measurements (see below) between the orbital evolution of Icarus and any candidate IFM. This procedure was successful in finding the dynamical relationship between (3200) Phaethon and (155140) 2005 UD (Ohtsuka et al. 2006, hereafter Paper I). No certain IFMs had been found in the Apollo asteroid database \footnote{e.g., http://cfa-www.harvard.edu/iau/lists/Apollosq.html} until very recently. However, we finally identified an extremely likely candidate from the latest MPECs (Minor Planet Electronic Circulars): a recently discovered Apollo asteroid 2007 $\mathrm{MK_6}$.

\section{Orbital integration of (1566) Icarus}

%% In a manner similar to \objectname authors can provide links to dataset
%% hosted at participating data centers via the \dataset{} command.  The
%% second curly bracket argument is printed in the text while the first
%% parentheses argument serves as the valid data set identifier.  Large
%% lists of data set are best provided in a table (see Table 3 for an example).
%% Valid data set identifiers should be obtained from the data center that
%% is currently hosting the data.
%%
%% Note that AASTeX interprets everything between the curly braces in the 
%% macro as regular text, so any special characters, e.g. "#" or "_," must be 
%% preceded by a backslash. Otherwise, you will get a LaTeX error when you 
%% compile your manuscript.  Special characters do not 
%% need to be escaped in the optional, square-bracket argument.

As preliminary work for the IFM survey and for measuring the time-lags (detailed in the next section) between Icarus and unknown potential IFMs, we calculated the orbital evolution of Icarus. We performed a backward and forward numerical integration of the KS (Kustaanheimo--Stiefel) regularized equation of motion (cf.\ Arakida \& Fukushima 2000, 2001), applying the 12th-order Adams method in double precision with a step size of 0.5 day. The integration covered 10000 BC to 10000 AD (JDT $-1931503.5$ to 5373520.5), and included terms of first order in the post-Newtonian approximation for the Sun's gravitational field since relativistic effects advance the line of apsides of Icarus at a rate of $10''$/century.

We also confirmed that the results of our numerical integration did not significantly change even when we adopted smaller step sizes or when we used other integration methods such as the extrapolation method. The initial orbital data of Icarus at osculation epoch 2007 Apr 10.0 TT = JDT 2454200.5 were taken from NASA JPL's HORIZONS System \footnote{http://ssd.jpl.nasa.gov/horizons.html} and are listed in Table \ref{tbl:orbits}. All the major planets from Mercury through Neptune and the quasi-planet, Pluto, were included as perturbing bodies (the Earth--Moon barycenter being one body, with the Moon's mass added to the Earth's). The coordinates of the major planets were taken from the JPL Planetary and Lunar Ephemeris DE408.

Over 20000 yr we found the orbital motion of Icarus to show a high degree of stability, with long-period secular changes according to the cycle in argument of perihelion $\omega$, also known as the Kozai cycle (Kozai 1962). The corresponding large-amplitude oscillations in $q$ and $i$, in antiphase with $e$, have period $\sim 25000$ yr, half that of the $\omega$ cycle. The $\omega$ period of $\sim 50000$ yr is somewhat larger than the $\sim 40000$ yr for Phaethon and 2005 UD (Paper I).

\section{Time lag $\Delta t$ of the orbital evolutions}

In the first stage of the formation of an asteroid family, the orbital energies ($\propto 1/a$) of bodies or fragments are slightly different from that of the precursor, since the motions of the released objects are slightly accelerated or decelerated relative to the precursor. This results in differences in their evolutionary rates under gravitational perturbations (there may additionally be differential nongravitational perturbations): then a time-lag (which hereafter we call $\Delta t$) in the orbital evolutions arises. At the starting epoch, $\Delta t \approx 0$ yr, and it tends to increase with time. We note that $\Delta t$ is not the time since separation, but rather quantifies how separated in phase two orbits have become in their respective (similar) secular perturbation cycles. For measuring a difference in the evolutionary phase of two orbits, $\Delta t$ is much more suitable than for example the difference in $\omega$, since (for highly eccentric orbits particularly) $d\omega/dt$ is strongly dependent on the phase within the Kozai cycle (cf.\ Fig. \ref{fig:orbits} later). Any IFM should show a very close orbital similarity with Icarus when shifted by the appropriate $\Delta t$ that brings both orbits to the same evolutionary phase.

The following successful studies applying this time-lag theory have been made so far: i) anticipation of the Marsden and Kracht comet groups' periodicity and their return: Ohtsuka et al. (2003) anticipated these comet groups, which initially had parabolic orbit solutions, as being fragments of Periodic Comet 96P/Machholz. Their prediction was shown to be correct when Sekanina \& Chodas (2005) linked orbits and found these comet groups to have orbital periods of 5--6 yr, corresponding to 96P's $\sim 5.2$ yr. The Marsden and Kracht comet groups thus turned out to be decameter-size members of the 96P--Quadrantid stream complex. ii) genetic relationship of Phaethon and 2005 UD: Paper I revealed 2005 UD as being the most likely large fragment of Phaethon. This dynamical relationship was confirmed by the physical studies of Jewitt \& Hsieh (2006) and Kinoshita et al. (2007), who classified both objects as F- or B-type. These taxonomic types are very rare, comprising only $\sim 5$\% of NEOs that have been classified; combined with the dynamical evidence, the genetic relationship of Phaethon and 2005 UD is beyond doubt.

This time-lag theory is straightforward and is now well established as a technique to demonstrate the existence of cometary stream complexes or likely NEO families; so it should be a useful tool to survey for IFMs.

\section{Survey}
\subsection{Procedure}

The survey for IFMs in the Apollo asteroid database and latest MPECs uses the same procedure as in Paper I. We again applied the following three criteria as the retrieving engine for our IFM survey. The first is the traditional orbital similarity criterion $D_{\rm SH}$ of Southworth \& Hawkins (1963), who defined $D_{\rm SH}$ as a distance between the orbits of two objects $A$ and $B$ in five-dimensional orbital element space $(e, q, \omega, \mathit{\Omega}, i)$, as follows:
\begin{eqnarray}
 D_{\rm SH}^{2} &=& \sum_{j=1}^{5} f_{j}^{2} \left( P_{A,j} - P_{B,j} \right)^2,
%%\label{equation-DSH}
\end{eqnarray}
where $P_{A \; \mathrm{or} \; B,j}$ are orbital elements and $f_j$ are functions of the elements that ensure suitable weights are given to each term in (1). Thus we searched for potential IFMs on the basis of Icarus' orbital evolution from the integration described in Section 2. For each Near-Earth Apollo, we found the minimum $D_{\rm SH}$ between it and Icarus, as Icarus' orbit evolves. A minimum $D_{\rm SH} \le 0.15$ means that Icarus and the given asteroid are within the probable association range.

The second and third criteria are the $C_1$ and $C_2$ integrals derived by Moiseev (1945) and Lidov (1961) respectively, which we calculate for candidates selected by $D_{\rm SH}$:
\begin{eqnarray}
 C_1 &=& \left( 1 - e^2 \right) \cos^2 i,\\
 C_2 &=& e^2 \left( 0.4 - \sin^2 i \: \sin^2 \omega \right) .
\end{eqnarray}
These integrals describe the secular orbital variations well. Both $C_1$ and $C_2$ are almost invariant for the orbital motions of Phaethon and 2005 UD (Paper I), and should also be useful criteria to distinguish IFMs.

\subsection{Detection of the IFM candidate: Near-Earth Apollo asteroid 2007 $\mathbf{MK_6}$}

In this way, we finally detected a very likely IFM candidate from the latest MPECs: Near-Earth Apollo asteroid 2007 $\mathrm{MK_6}$, which was recently discovered in the Catalina sky survey, on 2007 June 21.2 (Hill et al. 2007). Soon after, Ohtsuka (2007) made the identification of 2007 $\mathrm{MK_6}$ with another Apollo, 2006 $\mathrm{KT_{67}}$, so 2007 $\mathrm{MK_6}$ = 2006 $\mathrm{KT_{67}}$; the latter was both discovered (on 2006 May 26) and observed only (12 positions over a 1 day arc) by the Mt. Lemmon survey. This extended the arc to more than one year. Nakano successfully linked their orbits, based on 54 positions at two oppositions (covering 2006 May 26 to 2007 June 27) with an RMS residual of $0''.74$. The absolute magnitude $H \sim 19.9$ corresponds to an object a few hundred meters in size at most, if we assume 2007 $\mathrm{MK_6}$ is a high-albedo object such as an S-type.

\begin{figure}
\epsscale{1.0}
\plotone{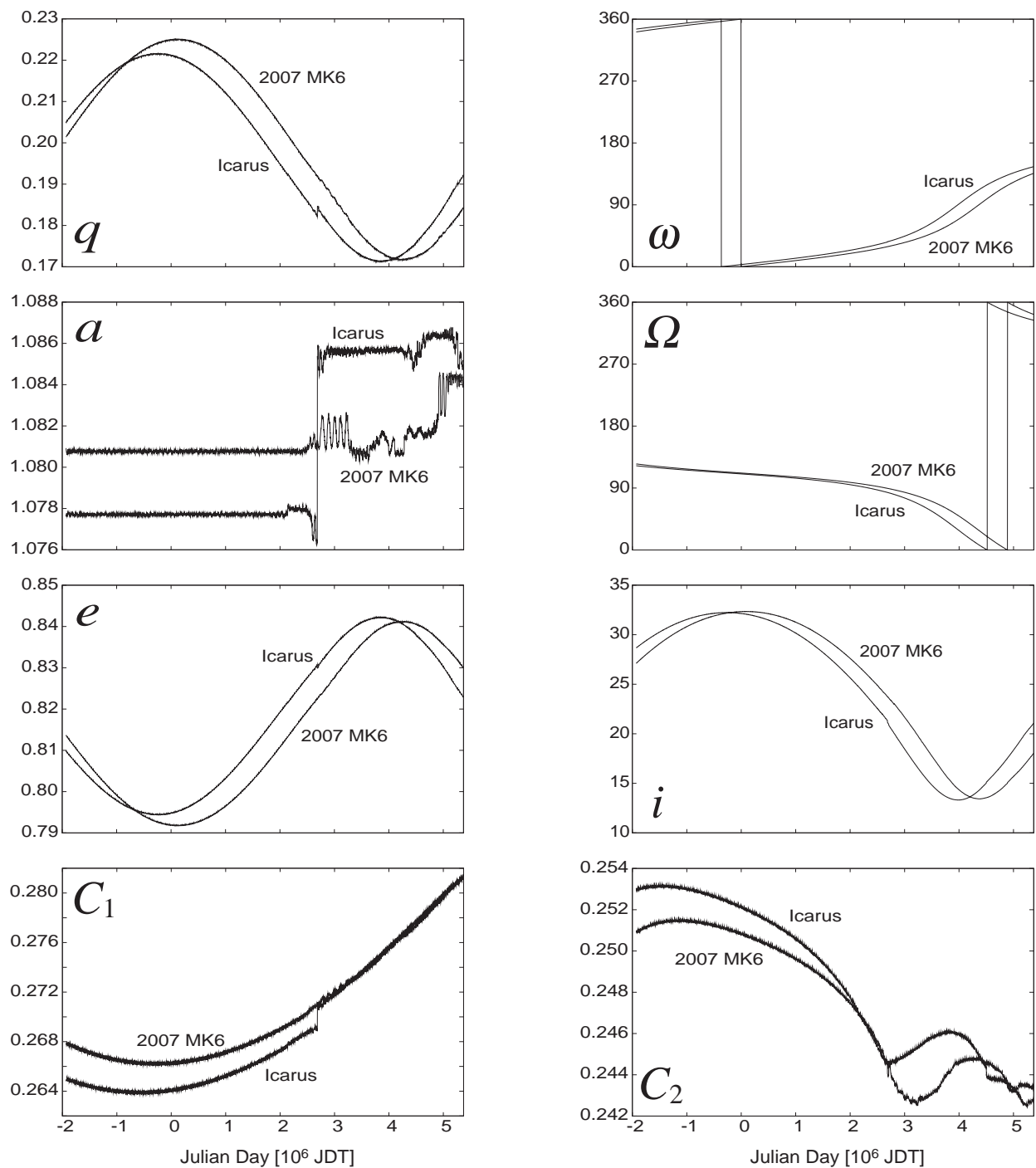}
\caption{Orbital evolution of (1566) Icarus and 2007 $\mathrm{MK_6}$.
    The eight graphs show: $q$ = perihelion distance in AU; $a$ = 
    semimajor axis in AU; $e$ = eccentricity;
    $\omega$ = argument of perihelion in degrees; $\mathit{\Omega}$ = 
    longitude of ascending node in degrees; $i$ = 
    inclination in degrees; and the $C_1$ and $C_2$ integrals.
    The abscissa of all is time (Julian Terrestrial Date, JDT).}
\label{fig:orbits}
\end{figure}

Using Nakano's data, listed in Table \ref{tbl:orbits}, we integrated 2007 $\mathrm{MK_6}$ using the same method as for Icarus. The dynamical evolutions of both asteroids are illustrated in Fig. \ref{fig:orbits}. Icarus and 2007 $\mathrm{MK_6}$ sometimes encounter the terrestrial planets. Encounters with Venus or Earth can cause changes in $a$ but these are small enough that the other elements display a stable secular evolution, as with Phaethon and 2005 UD (Paper I). Neither asteroid has a nodal intersection epoch with Venus or Earth in the past 10000 yr, hence the interval of constant $a$ in Fig. \ref{fig:orbits}.

Comparing the orbital elements of 2007 $\mathrm{MK_6}$ at the current epoch with the changing orbit of Icarus over time, as described in Section 4.1, we found a strikingly good match with Icarus at around 1034 AD (Table \ref{tbl:orbits}); thus $\Delta t \sim 1000$ yr. The corresponding minimum value of $D_{\rm SH}$ is only 0.0098. Both $\Delta t$ and $D_{\rm SH}$ are fairly small compared to the respective values $\sim 4600$ yr and 0.04 between Phaethon--2005 UD (Paper I). The $C_1$ and $C_2$ parameters are almost constant, within the ranges 0.26--0.28 and 0.24--0.25 respectively. Therefore 2007 $\mathrm{MK_6}$ is a very strong candidate IFM.

The two orbital evolutions show a similar profile, with quasi-sinusoidal changes, simply shifted by $\Delta t \sim 1000$ yr. Their smaller $\Delta t$ than Phaethon--2005 UD suggests a younger separation age, but $D_{\rm SH}$ between Icarus--2007 $\mathrm{MK_6}$ at the same osculation epoch has never been below 0.03 in our integration timespan. Only in quite rare cases (such as the Karin cluster in the main belt; Nesvorn\'y et al. 2002) can an exact separation age be found unambiguously, although we may certainly expect $D_{\rm SH}$ to have been smaller around the time that Icarus and 2007 $\mathrm{MK_6}$ separated. Some test integrations back $10^5$ yr tentatively show $\Delta t$ decreasing back in time but random small changes in $a$ due to close encounters make it hard to reach a precise quantitative conclusion about the separation age. However, this age is clearly within 10 Myr, the resurfacing age of Q-type NEOs, possibly two orders of magnitude shorter. The Icarus--2007 $\mathrm{MK_6}$ parent may well have been injected into the near-Earth environment of the order of 10 Myr ago (cf.\ Bottke et al. 2002), but with the separation occurring much more recently.

We also surveyed meteor data related to the IFMs. Consequently, we noticed a likely meteor swarm found by Sekanina (1973) in the Harvard (Havana) radar meteor orbit survey: the daytime Taurid-Perseid meteor swarm, recorded around June 18 in the radar's 1961--1965 term of operation. The orbital parameters are in good agreement with those of Icarus, as presented in Table \ref{tbl:IandT}. Their $D_{\rm SH} \sim 0.08$ is in the probable association range. Although we cannot accurately measure their $\Delta t$, both the current orbits look to be at almost the same evolutionary phase. No further orbital data were found in the radar meteor orbit database. We may therefore regard the Taurids-Perseids as a transient Earth-crossing IFM dust band rather than a cometary meteor stream.

\section{Discussion and Conclusions}

There have been numerous studies on the formation of main belt asteroid (MBA) families, and also some on NEO families. Statistical studies using NEO orbit data, based on orbital similarity, often generate positive results on the existence of NEO families. However, Fu et al. (2005) concluded that it is unlikely that these results are anything more than random fluctuations in the NEO orbit population. In a past IFM study, Steel et al. (1992) noted an orbital similarity between (5786) Talos = 1991 RC and Icarus. Their orbital elements, except for $\omega$ and $\mathit{\Omega}$, indeed coincide well with each other. However, Talos' longitude of perihelion remains widely separated (by $\sim 50^\circ$) from that of Icarus over the past 11000 yr integrated by Steel et al., so that it is difficult to verify a genetic relationship. If there exist NEO families having high-eccentricity and rather highly inclined orbits, then unless their origin is extremely recent, their differential orbital evolutions, shifted by $\Delta t$, will lead to their current orbital elements being drastically different. For this reason it is more complicated to search for NEO families than MBA families.

Nevertheless, we found Near-Earth Apollo asteroids Icarus and 2007 $\mathrm{MK_6}$ to be very likely candidates for IFMs, based on our time-lag theory. Their $\Delta t \sim 1000$ yr and minimized $D_{\rm SH} \sim 0.0098$ are even smaller than those, $\sim 4600$ yr and 0.04, of the well-established Phaethon--2005 UD relationship. Since Phaethon and 2005 UD may have a cometary origin (Paper I), therefore, the dynamical relationship between Icarus and 2007 $\mathrm{MK_6}$ along with a possible IFM dust band may constitute the first detection of an asteroidal NEO family, namely the ``Icarus asteroid family''. In this case, Icarus should be the parent body, but as it is only a 1-km size object, the Icarus family is on a smaller scale than MBA families.

The next Earth approaches of Icarus and 2007 $\mathrm{MK_6}$ will occur respectively on 2015 June 17 to 0.05 AU and 2016 June 15 to 0.10 AU, providing good opportunities to determine additional physical parameters and to further study their common origin. It is possible that further accurate astrometry and advances in the numerical analysis will eventually resolve the separation age.

\acknowledgments

The authors are grateful to the anonymous referee for his careful reading of the manuscript and for his comments.

\clearpage

%% the file, uncomment the \plotone, \plottwo, and
%% If you need a layout that cannot be achieved with \plotone or
%% \plottwo, you can invoke the graphicx package directly with the
%% \includegraphics command or use \plotfiddle. For more information,
%% please see the tutorial on "Using Electronic Art with AASTeX" in the
%% documentation section at the AASTeX Web site,
%% http://www.journals.uchicago.edu/AAS/AASTeX.
%%
%% The examples below also include sample markup for submission of
%% supplemental electronic materials. As always, be sure to check
%% the instructions to authors for the journal you are submitting to
%% for specific submissions guidelines as they vary from

%% This example uses \plotone to include an EPS file scaled to
%% 80% of its natural size with \epsscale. Its caption
%% has been written to indicate that additional figure parts will be
%% available in the electronic journal.

%\clearpage

\begin{table*}[htb]
\caption{Orbital parameters of (1566) Icarus and 2007 $\mathrm{MK_6}$ (equinox J2000)}
\label{tbl:orbits}
\centering
\begin{tabular}{lccc}
\hline
\hline
object & \multicolumn{2}{c}{(1566) Icarus} & 2007 $\mathrm{MK_6}$ \\%
\hline
osculation epoch (TT) & 2007 Apr. 10.0 & 1034 Jul. 03.0 & 2007 Apr. 10.0 \\
mean anomaly $M$      &  $285^\circ.14414$ & $161^\circ.45243$ & $336^\circ.75725$ \\
perihelion distance $q$ (AU) & 0.1866177 & 0.1930133 & 0.1959358 \\
semimajor axis $a$ (AU) & 1.0778849 &  1.0776443 & 1.0807494 \\
eccentricity $e$ & 0.8268668 & 0.8208933 & 0.8187038 \\
argument of perihelion $\omega$ & $31^\circ.29236$ & $25^\circ.37252$ & $25^\circ.38152$ \\
longitude of ascending node $\mathit{\Omega}$ & $88^\circ.08105$ & $93^\circ.55925$ & $92^\circ.94672$ \\
inclination $i$  & $22^\circ.85385$ & $25^\circ.07805$ & $25^\circ.15553$ \\
\# of observations &  711 & & 54 \\
arc (oppositions) & 1949--2006 (14) & & 2006--2007 (2) \\
RMS residual & $1''.12$ & & $0''.74$ \\
absolute mag. $H$  &  15.95 & & 19.9 \\
reference & JPL & this work &  Nakano \\
\hline
\end{tabular}
\end{table*}

\clearpage

\begin{table*}[htb]
\caption{(1566) Icarus and the Taurids-Perseids (Sekanina 1973) at almost the same evolutionary phase}
\label{tbl:IandT}
\centering
\begin{tabular}{lccccccc}
\hline
\hline
object & epoch & $q$ & $a$ & $e$ & $\omega$ & $\mathit{\Omega}$ & $i$ \\
       & (TT)  &  (AU) &  (AU) &     &     & (2000.0) &     \\
\hline
Icarus & 1963 Jul. 3.0 & 0.18697 & 1.07791 & 0.82655 & $31^\circ.032$ & $88^\circ.337$ & $22^\circ.940$ \\
Tau-Per & 1961--1965 & 0.163 & 1.268 & 0.871 & $36^\circ.3$ & $86^\circ.7$ & $23^\circ.3$ \\
\hline
\end{tabular}
\end{table*}

%% Any table notes must follow the \end{tabular} command.

%% If the table is more than one page long, the width of the table can vary
%% from page to page when the default \tablewidth is used, as below.  The
%% individual table widths for each page will be written to the log file; a
%% maximum tablewidth for the table can be computed from these values.
%% The \tablewidth argument can then be reset and the file reprocessed, so
%% that the table is of uniform width throughout. Try getting the widths
%% from the log file and changing the \tablewidth parameter to see how
%% adjusting this value affects table formatting.

%% The \dataset{} macro has also been applied to a few of the objects to
%% show how many observations can be tagged in a table.

\clearpage

%% Tables may also be prepared as separate files. See the accompanying
%% sample file table.tex for an example of an external table file.
%% To include an external file in your main document, use the \input
%% command. Uncomment the line below to include table.tex in this
%% sample file. (Note that you will need to comment out the \documentclass,
%% \begin{document}, and \end{document} commands from table.tex if you want
%% to include it in this document.)


%% The following command ends your manuscript. LaTeX will ignore any text

\end{document}